\documentclass[10pt, a4paper, oneside, leqno]{article}

\usepackage{amsmath}%
\usepackage{amsfonts}%
\usepackage{amssymb}%
\usepackage{graphicx}
\usepackage{here}
\usepackage{textcomp}
\usepackage{epstopdf}
\usepackage{color}
\usepackage{url}

\newcommand{\expectation}[1]{\mathbb{E}\left [ #1 \right]}
\newcommand{\cE}{\mathcal{E}}
\newcommand{\p}{\mathbb{P}}
\newcommand{\cL}{\mathcal{L}}

\begin{document}

\title{Quantifying Life Insurance Risk using Least-Squares Monte Carlo}
\author{C. Baumgart\footnote{Munich Re}, J. Krebs\footnote{Department of Mathematics, University of Kaiserslautern}, R. Lempertseder\footnote{Munich Re} and O. Pfaffel\footnote{Munich Re}}
\date{\today}
\maketitle


\section*{Executive Summary}
This article presents a stochastic framework to quantify the biometric risk of an insurance portfolio in solvency regimes such as Solvency II or the Swiss Solvency Test (SST). The main difficulty in this context constitutes in the proper representation of long term risks in the profit-loss distribution over a one year horizon. This will be resolved by using least-squares Monte Carlo methods to quantify the impact of new experience on the annual re-valuation of the portfolio. Therefore our stochastic model can be seen as an example for an internal model, as allowed under Solvency II or the SST. Since our model does not rely upon nested simulations it is computationally fast and easy to implement.

\paragraph{Outline of the article}
The first Section \ref{one year-horizon} introduces the notion of insurance risk in the context of an economic balance sheet. Section \ref{risk model} describes the stochastic simulation of forward projected paths used to model long-term insurance risks. The subsequent section \ref{LRFA} explains how Least-Squares Monte Carlo is applied to compute the Solvency Capital Requirement over a one year horizon. Section \ref{comparison} applies the theoretic framework to a reference mortality portfolio and compares the stochastic model with the Solvency II standard formula. In Section \ref{life_expectancy} we illustrate the impact of a 200-year event on the life expectancy of the portfolio. Finally, Section \ref{conclusion} concludes.

\newpage
\section{Introduction: measuring risk capital on a one year horizon} \label{one year-horizon}
Consider the simplified economic balance sheet\footnote{One may think of this as a simplification of the Solvency II balance sheet. However, the concepts of this article are general and can be applied to any economic solvency regime.} of an insurer in Table \ref{balance_sheet}. On the asset side we find the total value of all assets valued mark-to-market, $A_t$. The liability side consists of the own funds, $OF_t$, and the present value of all liabilities, $PV_t$\footnote{In Solvency II, the total liability (technical provisions) $TP_t$ is the sum of the best estimate liabilities $PV_t$ and a risk margin, thus $TP_t\neq PV_t$. Since the risk margin is, by concept, not at risk over one year, one still has $\Delta TP_t = \Delta PV_t$ so that all formulae in this article remain valid.}.
\begin{table}[H]
	\centering
		\begin{tabular}{ |c|c| }
		\hline
		\hline
				Assets & Liabilities \\
				\hline
				\hline
							& $OF_t$ \\
				$A_t$	&				\\
							&	$PV_t$		\\
				\hline
				\hline
		\end{tabular}
		\label{balance_sheet}
		\caption{The table depicts the simplified balance sheet of an insurer.}
\end{table}
We have the natural relation that $OF_t = A_t - PV_t$. Hence, we consider the own funds as the residual of assets and liabilities.
In general we define the risk that is associated with a random payoff $X$ as the deviation from its expectation, we write therefore $\Delta X := X - \expectation{X}$. We emphasize that there might be other understandings of risk in practice. With this definition we have the relation $\Delta OF_t = \Delta A_t - \Delta PV_t$.

Throughout this article we have the relation that $\Delta OF_t = - \Delta PV_t$ since we are only interested in biometric risk and as we assume that the asset side is not subject to biometric risk. In general $\Delta PV_t$ will be a function of various inputs, for instance, mortality rates, lapse rates, stock prices and  yield curves. Since we investigate biometric risk we assume inputs which stem from the capital market side to be deterministic, e.g. we take yield curves as deterministic.

We are interested in the amount of capital the insurer has to hold in order to guarantee solvency over a one year horizon. As a risk measure, we use the Value-at-Risk of the own funds subject to a confidence level of 99.5\% over a one-year period. This is in line with Solvency II: the \textit{Solvency Capital Requirement} (SCR) is given by $SCR = VaR_{0.005}(\Delta OF_1) = VaR_{0.995} (\Delta PV_1)$. Nevertheless, the theory developed in this article can be applied to any risk measure that is a function of the own funds distribution over one year, such as the Tail-value-at-Risk used for the SST, for example.

\section{The Stochastic Risk Model}\label{risk model}

\paragraph{Stochastic adjustment of projected cash flows}
In this section we introduce a stochastic model that produces simulated paths for the development of future cash flows. Without loss of generality we consider individual life (mortality) business rather than longevity business and consider future mortality and lapse rates as major risk drivers.
For net cash flows of a single portfolio we immediately see that certain components are risky whereas others are not, or rather, not significantly risky. For instance, we may assume the cash flow of premiums to behave as predicted by the underwriter whereas the projection of death rates carries uncertainty. To describe this we attribute to each risky part of the net cash flow an individual factor that adjusts for its risk. If we assume that the net cash flow which accrues in each year decomposes\footnote{To increase readability, we deduct $Expenses$, $Commissions$ and $Taxes$ from the premiums a priori.} into
\[
		NCF(t) = NetPremium(t) - Death Payments(t),
\]
then the corresponding stochastic model is given by
\[
		NCF(t) = V(t)\cdot\left( NetPremium(t) - AF(t) \cdot Death Payments(t)\right)
\]
with a suitably chosen adjustment factor $AF$. Figure \ref{fig:funnel_mortality2} (later in this article) illustrates sample paths and the corresponding quantiles for the adjustment factor used to model mortality risk in a reference mortality portfolio.
The process $V$ is an adjustment factor for the portfolio volume. This modification is necessary as the portfolio volume is now stochastic due to stochastic mortality rates. 

We identify for each adjustment factor various sources of uncertainty: the adjustment factor for death payments should capture the risk of mis-estimation of future mortality trends (\textit{trend risk}) as well as the risk of short term catastrophic events (\textit{calamity risk}). Moreover, when considering a portfolio of insureds, $AF$ should capture the risk of a mismatch between own portfolio experience and the overall population (\textit{basis risk}). In the upcoming Section \ref{comparison} we assess each of these risk sources for a reference mortality portfolio.

\paragraph{Computation of solvency capital over a one year period}
In order to guarantee solvency of the insurer over a one year horizon, we are interested in the change of the present value of net cash flows over one year. The present value of net cash flows is given by	$PV = \sum_{t=1}^{\mathcal{T}} \delta(t) NCF(t)$, where $\delta(t)$ is the discount factor for the net cash flow that accrues in period $t$. 

Hence, from the perspective $t=0$ we have
\[
		\Delta PV = \sum_{t=1}^{\mathcal{T}} \delta(t) NCF(t) - \expectation{\sum_{t=1}^{\mathcal{T}} \delta(t) NCF(t)}.
\]
In other words, $\Delta PV$ is the present value of the difference between the actual net cash flows and their best estimate which is derived given all information available at time $t=0$. To determine the $SCR$, we have to compute the re-valuation of this random variable given the information update in $t=1$, i.e., after observing the impact of the realization of the risk drivers on the portfolio of insureds\footnote{Of course, this does not imply that the realizations of all risk drivers are observed directly. Mathematically, one has to account for this by choosing an appropriate filtration of sigma-algebras.}. This yields the following equation
\begin{align}
	&\expectation{\Delta(PV)\,\Big|\, AF(1) } \nonumber \\
	&= \expectation{ \sum_{t=1}^{\mathcal{T}} \delta(t) NCF(t) \,\Big|\,AF(1)} - \expectation{\sum_{t=1}^{\mathcal{T}} \delta(t) NCF(t) } \nonumber \\
	&= \delta(1) \left\{ NCF(1)-\expectation{ NCF(1)  } \right\}  + ... \label{ev} \\
	&\quad ... + \sum_{t=2}^{\mathcal{T}}  \delta(t) \left\{ \expectation{ NCF(t) \,\Big|\,AF(1) } - \expectation{ NCF(t) }		\right \}. \label{ac}
\end{align}
To keep the notation simple, we do not enumerate explicitly each risk driver in the conditional expectation, but rather use the symbol $AF(1)$ to denote the observed impact of all risk drivers. In the upcoming section we present a method to compute the conditional expectation in equation \eqref{ac} using a \textit{Least-Squares Monte Carlo} (LSMC) approach

\section{Computation of solvency capital using least-square Monte Carlo methods}\label{LRFA}

In the following we present an approach that allows to derive the distribution of $\expectation{ \Delta(PV) \,|\, AF(1) }$ with Monte Carlo methods. In practice this can be considered as the valuation update based on the transition from period $t=0$ to period $t=1$. We call equation \eqref{ev} the \textit{experience variance} and equation \eqref{ac} the \textit{assumption change}. The first term clarifies the \textit{variance} of the actual net cash flow in period 1 and its estimate in period 0 that is \textit{experienced} during the transition from period 0 to period 1. Furthermore, this update in information entails that all future net cash flows have to be \textit{changed} accordingly based on new \textit{assumptions} made in period 1, i.e. the conditional expectation in \eqref{ac} has to be computed. Note that the assumption change entirely relies on the experience variance in the previous year and not on any other (external) information. This means that the portfolio has to be sufficiently large so that one can assign full credibility to one year portfolio experience. \\ 

In his re-valuation process, an actuary would use the information update after one year for a re-calibration of the biometric assumptions and use them for an updated forward projection of net cash flows. We omit the intermediate re-calibration step since we are only interested in the change of the present value of the updated forward projection after one year. The reader can find an approach which does incorporate the idea of re-calibration in the context of longevity risk in \cite{DHR}, \cite{PS} and \cite{InsuranceRisk}. B{\"o}rger et al. \cite{borger2011modeling} are using generalized linear models for a re-assessment of mortality trend after one year. An alternative approach constitutes in using replicating portfolios (cf. \cite{pelsser2015difference}), or nested simulations (cf. \cite{BBR}).
The LSMC methods in the context of finance date back at least to Longstaff and Schwarz who used it in 2001 to value American options \cite{LS}. In the context of financial portfolio analysis, LSMC is analyzed in depth by \cite{danielsson2014estimation}. There are various variants of LSMC methods which are used to determine the risk associated with insurance portfolios. 
Current literature mostly focuses on market risks and treats biometric risks in a simplified way, e.g. using deterministic stress scenarios: H{\"o}rig and Leitschkis \cite{Milliman} introduce a procedure that combines nested simulations with linear regression. Krah et al. \cite{krah2018least} and Floryszczak et al. \cite{floryszczak2016inside} focus on deterministic stress scenarios for biometric risks (e.g. mortality) and model market / credit risk stochastically. Peters et al. \cite{peters2017bayesian} provide a Bayesian framework with a focus on coherent capital allocation that is compliant with the Swiss Solvency Test. Ha and Bauer \cite{ha2015least} frame the estimation problem via a loss operator that maps future payoffs to the conditional expected value at the risk horizon to obtain an optimal choice for the basis functions in the regression. In \cite{ha2015least} asset price, short rate and force of mortality are jointly modelled to price a guaranteed minimum withdrawal benefit using LSMC.\\

Our idea is to approximate equation \eqref{ac} with a multilinear function as
\begin{align}\begin{split}\label{lrfa2}
	& \sum_{t=2}^{\mathcal{T}}  \delta(t) \left\{ \expectation{ NCF(t) \,|\, AF(1) } - \expectation{ NCF(t)}		\right \} \\
	& \approx \sum_{i} \alpha^*_{\text{risk driver}_i} \left({AF}^{\text{risk driver}_i}(1)-1\right),	
\end{split}\end{align}
where $risk driver_i$ constitute various risk drivers affecting the portfolio of insureds, such as the mortality rates for different age groups or lapse rates. Note that the model needs no intercept, since the biometric assumptions in $t=0$ are best estimates and therefore $\expectation(AF^{\text{risk driver}_i}(1))=1$.

In Equation \eqref{lrfa2} we postulate that $\sum_{t=2}^{\mathcal{T}}  \delta(t) (NCF(t) - \expectation{ NCF(t)})$ equals the linear function above plus an error term whose conditional expectation vanishes. In subsequent sections we plot this relationship for two sample portfolios in Figures \ref{fig:MP0001_RiskDriver_vs_Risk} and  \ref{Risk_vs_RiskDriver_3d}. The key idea is to approximate this unconditional random variable by simulations. We denote the left hand side of \eqref{lrfa2} by $PV(\Delta NCF(2:{\mathcal{T}}))$ and run $N$ (say, 100,000) simulations of the risk drivers to compute projected net cash flows. 

Subsequently, we regress the variable $\sum_{t=2}^{\mathcal{T}}  \delta(t) NCF(t) - \expectation{ NCF(t)}$ on the (observable impact of the) realizations of the risk drivers of the first year in order to obtain estimates for the coefficients $\alpha^*_{\text{risk driver}_i}$. By virtue of equations \eqref{ev}, \eqref{ac} and \eqref{lrfa2} we arrive at the following approximation
\begin{align}
	&\expectation{\Delta(PV)\,|\, AF(1) } \nonumber \\
	&\approx  \delta(1) \left\{ NCF(1)-\expectation{ NCF(1) } \right\}  + \sum_{i} \alpha^*_{\text{risk driver}_i} \left({AF}^{\text{risk driver}_i}(1)-1\right) \label{eq_pl_dist}
\end{align}
If we insert the simulated values for the risk drivers in the above equation we obtain the distribution of $\expectation{ \Delta(PV)\,|\, AF(1)}$. The $99.5$ percentile of this distribution yields the $\text{VaR}_{0.995}$,
\begin{align}
& \text{VaR}_{0.995} \\
&= q_{99.5\%}\left( \delta(1) \left\{ NCF(1)-\expectation{ NCF(1) } \right\}  + \sum_{i} \alpha^*_{\text{risk driver}_i} \left({AF}^{\text{risk driver}_i}(1)-1\right) \right) \nonumber
\end{align}
Any other quantiles, or functions of quantiles (such as the Expected Shortfall), can be computed likewise. Additionally, we obtain more than just the risk distribution: the coefficients $\alpha^*_{\text{risk driver}_i}$ of the linear regression show us the impact of each risk driver on an assumption change given a particular biometric first year realization, cf. Figure \ref{fig:MP0001_RiskDriver_vs_Risk} later in this text.

\section{Calculation of the Solvency Capital Requirement of Solvency II}\label{comparison}

\subsection{A short review of the Capital Requirement via the Standard Formula}
One main difference between our presented stochastic model and the standard formula is that the latter does not simulate paths of a stochastic process but instead assumes a deterministic stress scenario. We consider the following stress scenarios that are relevant for mortality business:\\
\\
\begin{tabular}[H]{||p{3.9cm}|p{3.5cm}|p{3.3cm}||}  
 \hline \hline Scenario Name & Scenario Description & Affected Risk Driver\\
 \hline
 \hline 
  Lapse shock (\#1) & Lapse shock of 40 \% of underlying portfolio & 40 \% x SaR (Sum at Risk) lapse shock in year one \\ \hline
  Lapse increase (\#2) & 50 \% increase in lapse rates compared to b.e. over projection period & + 50 \% lapse rates \\ \hline
  Lapse decrease (\#3) & 50 \% decrease in lapse rates compared to b.e. over projection period & - 50 \% lapse rates \\ \hline
  Mortality increase (\#4) & 15 \% increase in mortality rates compared to b.e. over projection period & + 15 \% mortality rates \\ \hline
  Life catastrophe (\#6) & Standard Approach Cat Risk & +1.5 \textperthousand $ $ x SaR one-year shock\\ \hline \hline
 \end{tabular}\\ \\
Denote by $PVoFP_{det}$ the present value of future profits in the deterministic scenario, i.e. the scenario that relies on the best estimate, and abbreviate by $PVoFP_{k}$ the present value of future profits from scenario number $k$. Then the SCR according to the standard formula, which we denote by $\text{VaR}_{0.995}^{SF}$, is computed as follows: First we calculate the sub solvency capital requirements
\begin{align*}
	VaR_{0.995,\text{lapse}}(SF) &:= \max( PVoFP_{det} - PVNCF_{\#1},PVoFP_{det} - PVNCF_{\#2},\ldots \\
	&\quad \ldots, PVoFP_{det} - PVNCF_{\#3},0), \\
	VaR_{0.995,\text{mortality}}(SF) &:= \max(PVoFP_{det} - PVNCF_{\#4},0), \\
	VaR_{0.995,\text{life catastrophe}}(SF) &:= \max(PVoFP_{det} - PVNCF_{\#6},0).
\end{align*}	
And in the second step we aggregate them by means of the \textit{square root formula} 
\begin{align*}
VaR_{0.995}(SF) &:= \Big( VaR_{0.995,\text{lapse}}^2 + VaR_{0.995,\text{mortality}}^2 
+ VaR_{0.995,\text{life catastrophe}}^2 + \ldots \\
&\quad \ldots + 2\cdot \frac{1}{4} \cdot VaR_{0.995,\text{mortality}}\cdot VaR_{0.995,\text{life catastrophe}} + \ldots \\
&\quad \ldots + 2\cdot \frac{1}{4} \cdot VaR_{0.995,\text{lapse}}\cdot VaR_{0.995,\text{life catastrophe}} 
\Big)^{\frac{1}{2}}
\end{align*}
The standard formula is intended to compute the Value-at-Risk subject to a level of $99.5\%$. Whereas our presented stochastic model can be calibrated to capture the specific risk profile of a portfolio, the standard formula has to be applied in the same way regardless of individual aspects such as the size and riskiness of a portfolio. One immediately recognizes that this lack of discrimination can be a serious drawback.


\subsection{SCR of a sample mortality portfolio}

In this section we consider a sample portfolio $M$ which comprises of whole life insurance contracts - real, not simulated data. We compare the SCR for this portfolio calculated with the introduced stochastic model and with the standard formula. The primary focus of this section is on the methodology: the main steps necessary to build up a stochastic model that can be used with LSMC. Our parameterisation of the stochastic model is motivated in the text below and broadly in line with the authors' own experience. Nevertheless, we need to stress that our parametrization is only exemplary for a practical comparison between our stochastic model with a model based on deterministic stresses. In particular, it should not be seen as a basis for any real risk assessment. We begin by explaining the specification of the stochastic model for $M$.

\subsubsection{Stochastic risk model for population mortality} 

\begin{figure}[H]
	\centering	\includegraphics[width=.84\textwidth]{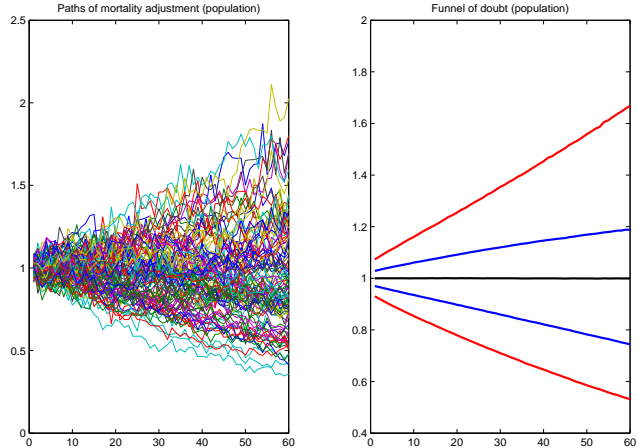}
	\caption{The figure on the left depicts 100 sample paths for the mortality adjustment relative to the best estimate. The figure on the right shows the corresponding funnel of doubt. The red lines are the estimated 95 \% (resp. 5 \%) quantiles, the lines in blue the 75 \% (resp. 25 \%) quantiles. The mean is plotted in black. This graphic is based on 100,000 simulations.}
	\label{fig:funnel_mortality}
\end{figure}

We included the following risk drivers in the model: base mortality\footnote{Briefly just referred to mortality in the following.}, pandemic mortality and lapse rates. For the considered portfolio base mortality is the main risk driver\footnote{This will be seen in Figure \ref{fig:MP0001_RiskDriver_vs_Risk}.} and therefore receives the main attention\footnote{Future lapse rates are simulated using a Brownian motion, pandemic mortality is given via a Pareto distribution with its parameters chosen such that the 98 \% quantile corresponds to an excess mortality of 0.4 \textperthousand \, and the 99.9 \% quantile to 5 \textperthousand; this parametrization is based on a recommendation of the American Society of Actuaries, (cf. \cite{USA}). There exist other recommendations, for instance \cite{Moodys}.}. In order to specify an adequate model for mortality risk, a Bayesian Lee Carter (BLC) model is fitted to UK mortality rates from 1956 to 2013; a similar analysis has also been performed by \cite{pedroza2006bayesian}. The trend risk model is then based on the volatility of the BLC around its mean, cf. \cite{schiller2011modelling} for details, and further smoothing adjustments. Since none of the two aforementioned papers states the full parameterization of the BLC model, we have no parameters we can directly compare our calibration with. However, we see from a study of \cite{cairns2011mortality} based on England \& Wales (male) mortality data, that, for the classical Lee Carter model, an increase in mortality rates of roughly 45\% relative to the best estimate over 40 years corresponds to the upper 5\% quantile (cf. Figure 3 of \cite{cairns2011mortality}). We observe that our parameterization is in the same range. Figure \ref{fig:funnel_mortality} shows the sample paths of the adjustment factor for mortality trend risk and the resulting funnel of doubt: the upper red line corresponds to the upper 5\% quantile and lies at roughly $1.45$ after $40$ years; since $1$ corresponds to the best estimate, we have an increase of 45\%.



\paragraph{A first SCR in the stochastic model considering only population mortality risk.}
The least-squares Monte Carlo model with above parameterization yields an SCR for portfolio $M$ of $\text{VaR}_{0.995}(M) = \$1,418m$. Here $\text{VaR}_{0.995}(M)$ equals the 0.995 quantile of the distribution that is generated by equation \eqref{eq_pl_dist} based on the regression performed on the simulated net cash flows of $M$.

\paragraph{SCR via the standard formula.}
The SCR of portfolio $M$ computed with the standard formula amounts to $\text{VaR}_{0.995}^{SF}(M) =\$4,426m$. This means the Solvency Capital Requirement computed with the standard formula is more than three times higher than the capital requirement based on the stochastic model that allows for individualities of the portfolio $M$. This difference may be largely attributed to 
\begin{enumerate}
	\item the fact that the stochastic model so far only considered population mortality risk rather than \textbf{portfolio} mortality, and
	\item an implicit assumption on the degree of \textbf{diversification} in the portfolio that is underlying the standard formula calibration.
\end{enumerate}
We'll cover both points in the next section.

\subsubsection{Stochastic risk model for portfolio mortality} 
Even in a large portfolio the insureds will not represent the underlying population with respect to age structure, geographic location or socio-economic class. We defined basis risk as the risk of a mismatch between portfolio and population mortality. It has been well known for decades that smoking habits significantly impact mortality rates (cf. \cite{lew1987differences}, for instance), therefore insureds are being discriminated by smoking status: smokers pay higher premiums than non-smokers. 
It would be an interesting aspect of its own to analyze which factors that impact future mortality are currently unknown to insurers but might affect insured portfolios differently than the overall population. For example, let us assume that higher mobile phone usage triggers a higher risk of developing cancer in future years. Via adverse portfolio selection our portfolio could contain a higher ratio of intensive mobile phone users than one finds in the population (this could also be linked to the socio-economic distribution of our portfolio diverging from the population). Since mobile phone usage is typically not known to an insurer, this discrepancy may increase future portfolio mortality while the overall population mortality is significantly less affected.\\
In order to keep this exercise short we will not analyze unknown drivers of basis risk but, as a proof of principle, make the assumption that the smoking habits of our insureds are unknown. This way we can give an assessment of basis risk that uses generally established and accepted data. Moreover, 
this is not an unrealistic assumption in case of a mortality analysis for disability or income protection business, where smoker information is hardly available, in particular for older in-force business. 

\paragraph{Example for a consideration of basis risk.}  
 The `00' series base mortality tables of CMI (Continuous Mortality Investigation) are based on 1999-2002 experience collected from UK insurance companies and publicly available on CMI's website \cite{CMItables}. Since we are considering an in-force portfolio consisting of whole life insurance policies we are referring to the tables for permanent assurances with a select duration of two or more years. For males with an age between 35 and 65 years, smoker mortality is 1.7 to 2.2 times higher than non-smoker mortality, for females this factor varies from 1.3 to 2.5. In average, the ratio of smoker to non-smoker mortality is roughly 2 for males as well as females. Currently, 35\% of the UK population are smokers. Given the size of portfolio M, we may assume that this share ranges from 31\% to 39\% in our portfolio. Denote by $q_p(\alpha)$ the portfolio mortality given that the fraction of smokers relative to all insureds is equal to $\alpha$. The unfavourable case is when $\alpha$ is larger than 35\%. We calibrate the standard deviation of the basis risk component such that
\[ 1 + \sigma_\mathrm{basis risk} = \frac{q_p(39\%)}{q_p(35\%)}. \]
We refer to $M^\mathrm{ref}$ as the portfolio $M$ according to this basis risk parameterization. In this setup, the LSMC model yields $\text{VaR}_{0.995}(M^\mathrm{ref}) = \$2,388m$. Even though this risk charge is 68\% higher than the one computed without basis risk, this means point 1 mentioned above only partly explains the difference to the standard formula. Based on an SCR of  $\text{VaR}_{0.995}(M^\mathrm{ref}) = \$2,388m$, we see that the SCR computed via the standard formula is still another 85\% higher. This may be attributed to a large degree due to point 2, as the following example shows.

\paragraph{Impact of diversification.}  
When we consider a sub-portfolio $M^\mathrm{sub}$ of $M^\mathrm{ref}$, basis risk increases since it is more likely that a small portfolio is even less represented by the overall population than a large portfolio. It is reasonable\footnote{$M^\mathrm{sub}$ is much smaller than portfolio $M^{ref}$. For instance, the PVoFP of $M^\mathrm{sub}$ is equal to \$23.5m, where the PVoFP of $M^\mathrm{ref}$ is equal to \$262.0m} that the share of smokers in portfolio $M^\mathrm{sub}$ lies between 27\% and 43\%. With the same reasoning as above for portfolio $M^\mathrm{ref}$, this increases basis risk and yields $\text{VaR}_{0.995}(M^\mathrm{sub}) = \$382m$. The standard formula yields $\text{VaR}_{0.995}^{SF} (M^\mathrm{sub}) = \$397m$. Thus we see that for our sub-portfolio, the SCR computed with the stochastic model is in the same range as the SCR computed via the standard formula. This indicates that the biometric stresses of the standard formula may be adequate for an insurance portfolio that is subject to a high basis risk due to rather low diversification, but seems too conservative for a bigger and well diversified portfolio such as $M^\mathrm{ref}$. Note that a related analysis of \cite{borger2010deterministic} produced similar results.\\

Finally, both examples show that our stochastic risk model offers enough flexibility to satisfy the needs of a large insurer with a well-diversified portfolio as well as the needs of a smaller insurer that is subject to a much higher basis risk.

\subsubsection{Discussion of the results and properties of the stochastic model}
\begin{figure}[H]
	\centering	\includegraphics[width=.84\textwidth]{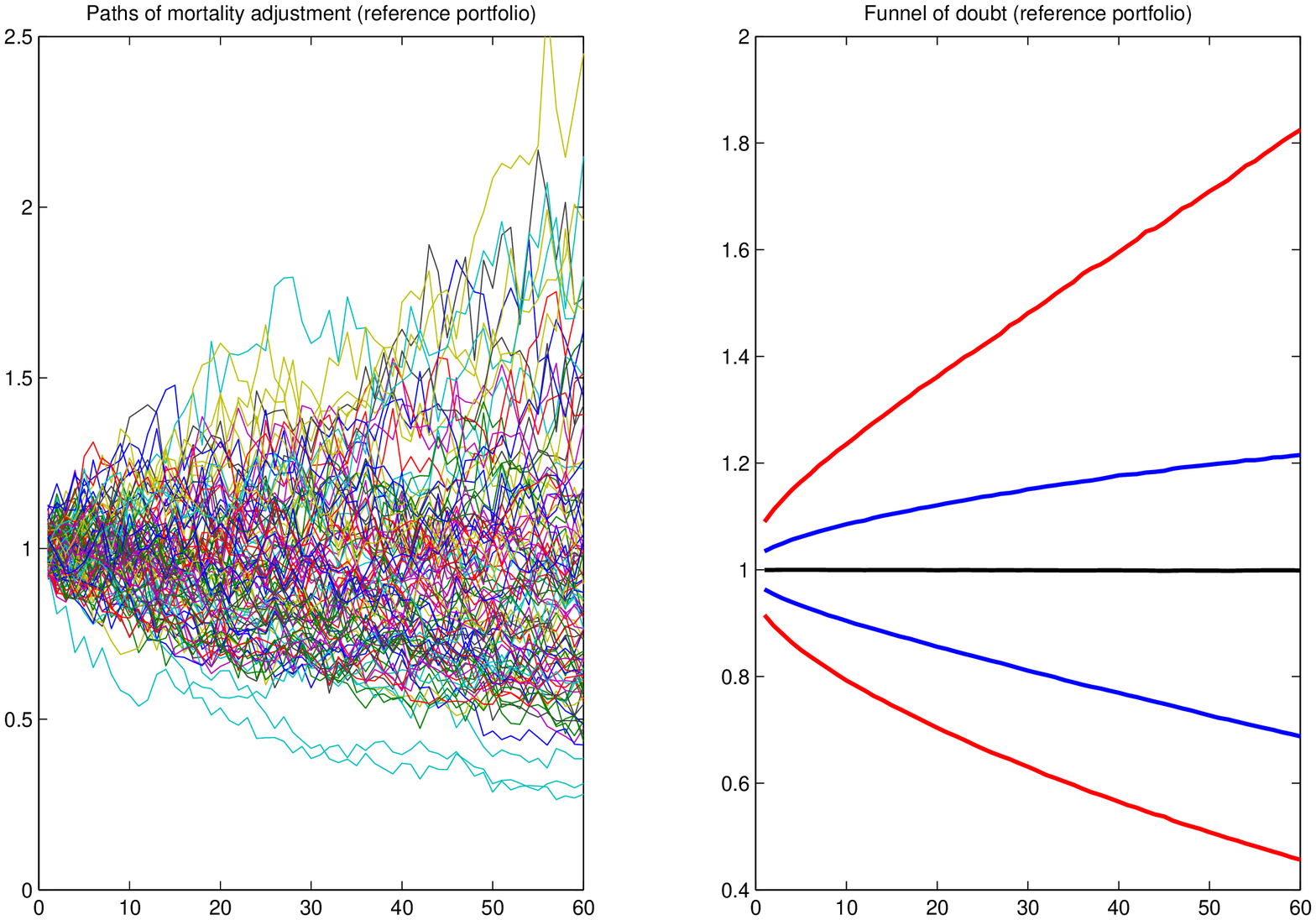}
	\caption{100 sample paths for the mortality adjustment and the corresponding funnel of doubt. The red lines are the estimated 95 \% (resp. 5 \%) quantiles, the lines in blue the 75 \% (resp. 25 \%) quantiles. The mean is plotted in black.}\label{fig:funnel_mortality2}
\end{figure}

\begin{figure}[H]
	\centering	\includegraphics[width=.84\textwidth]{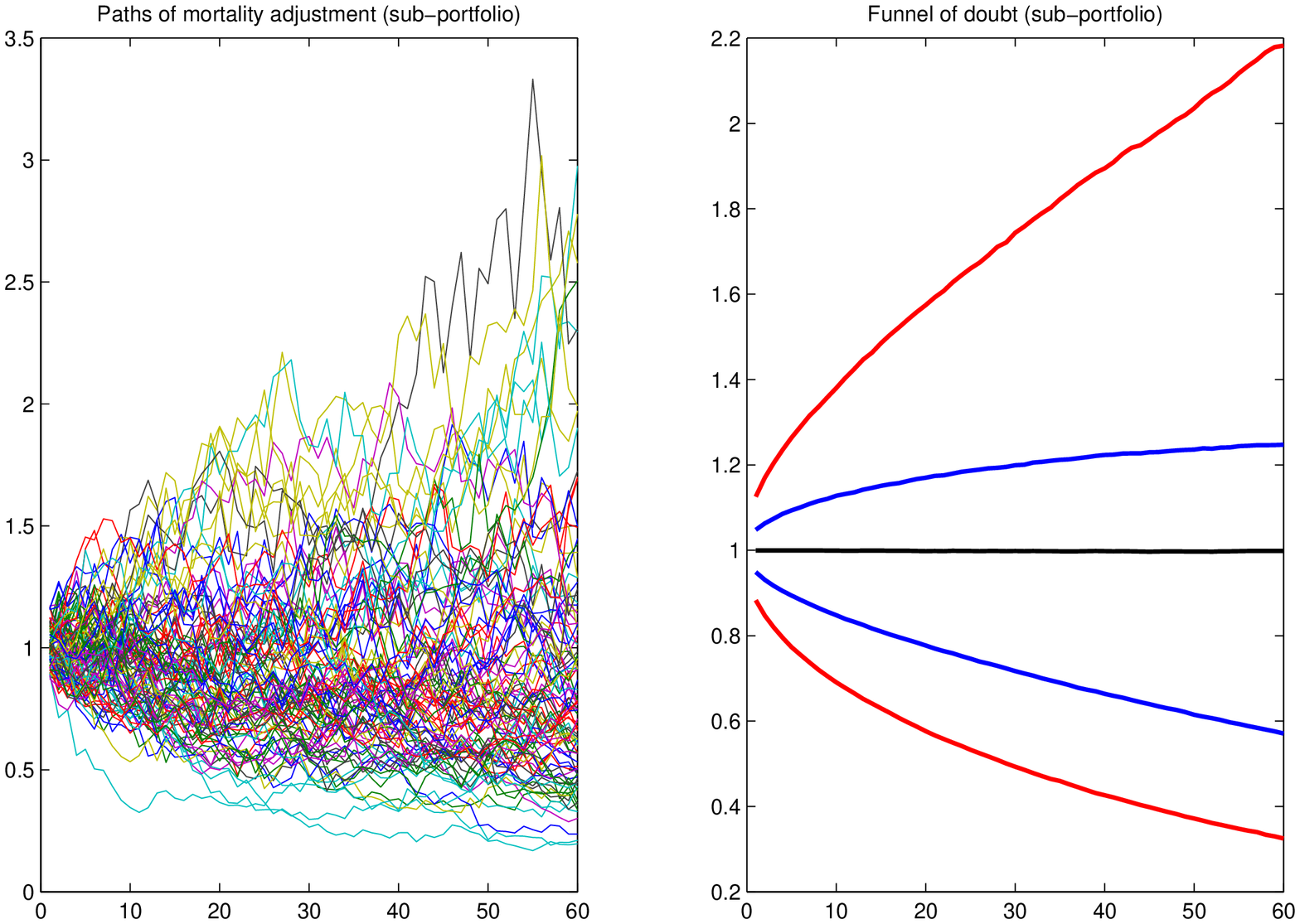}
\caption{100 sample paths for the mortality adjustment and the corresponding funnel of doubt. The red lines are the estimated 95 \% (resp. 5 \%) quantiles, the lines in blue the 75 \% (resp. 25 \%) quantiles. The mean is plotted in black.}\label{fig:funnel_mortality3}	
\end{figure}

Figure \ref{fig:funnel_mortality2} and \ref{fig:funnel_mortality3} show sample paths and the funnel of doubt of $M^\mathrm{ref}$ and $M^\mathrm{sub}$ for mortality risk. One immediately sees that the paths in both plots are more spread out than the population mortality paths in Figure \ref{fig:funnel_mortality}, and that the paths of the sub-portfolio are again significantly more spread out than the one for the reference portfolio. Consistently to these results, Plat \cite{plat2009stochastic} has analyzed basis risk of two example insurance portfolios with the result that the funnel of doubt of the medium-sized portfolio significantly exceeds the one of the large portfolio.

\begin{figure}[H]
	\centering	\includegraphics[width=.77\textwidth]{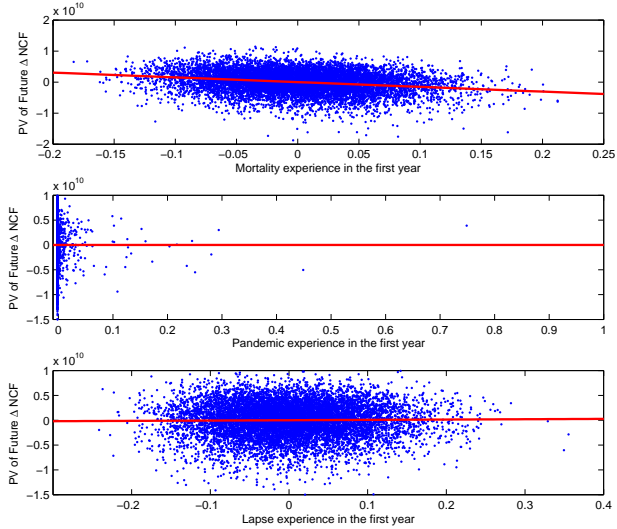}
	\caption{Dependence of the present value of future $\Delta$ net cash flows on the risk drivers. The red lines show the regression output projected from the three dimensional hyperplane to the one dimensional line with respect to the level where the other risk factors are zero. The figure is based on 100,000 simulations.}
	\label{fig:MP0001_RiskDriver_vs_Risk}
\end{figure}

\begin{figure}[H]
	\centering	\includegraphics[width=.77\textwidth]{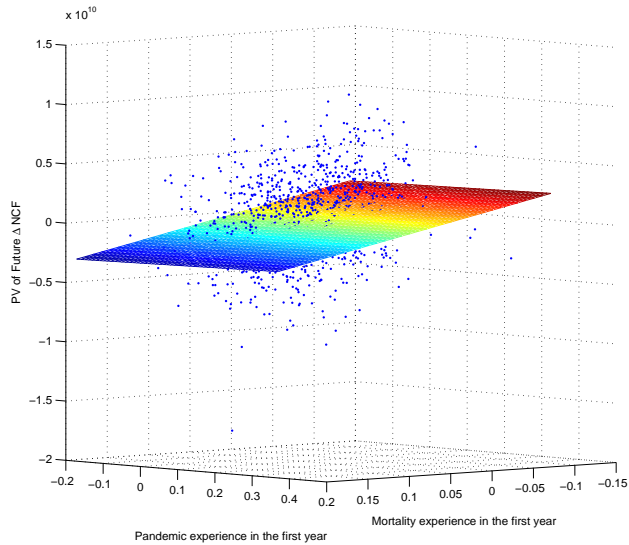}
	\caption{Dependence of the present value of future $\Delta$ net cash flows on the two risk drivers mortality and lapse, including the regression surface. The figure is based on 100,000 simulations.}
	\label{Risk_vs_RiskDriver_3d}
\end{figure}

In Figure \ref{fig:MP0001_RiskDriver_vs_Risk} the risk of the portfolio $M$ is plotted against some of its regressors. One recognizes the negative impact of higher mortality rates on future profits; this is because an increase in observed mortality rates in the first year presumably leads to higher payments in the immediate future and, thus, reduces profits. Moreover, one sees that a higher mortality experience after one year has as much stronger impact on the assumption change than, for example, a higher experience in lapse rates. Figure \ref{Risk_vs_RiskDriver_3d} depicts the dependence of the assumption change on the two risk drivers mortality and lapse.

\paragraph{A note on the $R^2$ of the regression and the linear approximation.}
In statistical data analysis, a low $R^2$ shows that the regression model explains only a fraction of the variability in the data, thus indicating a poor fit of the model. This then motivates the use of additional regressors or more complex basis functions. Here the situation is quite different: the low $R^2$ of the least-squares linear regression is naturally linked to the fact that the net cash flows are projected over a long time horizon and thus vary considerably. Hence, the conditional expectation of $\Delta(PV)$ after one year of experience still has a high standard deviation, since one can explain the outcome of the next (say) 60 years of future net cash flows based on the observations of a single year only to a small amount. Longstaff \& Schwartz \cite{LS} remark this point as well.\\
Nevertheless, as an additional validation, we included quadratic terms in the regression model. The result was a not significantly lower $\text{VaR}_{0.995}$, thus it is not advisable to add these additional terms as this would increase model risk but not provide a statistically better result.


\section{Plausibility check: re-estimating life expectancy after one year}\label{life_expectancy}
In the last section we have seen that the Value-at-Risk of both sample portfolios computed with the stochastic model is higher than the Value-at-Risk computed with the standard formula. In this section we want to illustrate this difference by looking at the impact of both models on the remaining life expectancy of the portfolio, a quantity that is more illustrative than a capital requirement measured in US\$.

To simplify the analysis we consider aggregated mortality rates of the entire portfolio. In particular, the is no distinction between gender or age. The forward looking mortality rates $q_1,q_2,\ldots,q_{\mathcal{T}}$ for the time horizon $[0,\mathcal{T}]$ can be interpreted as follows: Provided an  insured has survived until period $k-1$, the probability of surviving the time interval $(k-1,k]$ is $1-q_k$. Denote by $\cL$ the remaining lifetime of our portfolio from the perspective $t=0$; then
\[
		\cL \ge k \text{ with probability } (1-q_1)\cdot(1-q_2)\cdot \ldots \cdot (1-q_k).
\]
The remaining life expectancy of the portfolio with respect to the time horizon $\mathcal{T}$ amounts to
\begin{align*}
		\cE := \expectation{\cL} = \int_0^{\mathcal{T}}  \p( \cL \ge t)\,\mathrm{d}t = \sum_{k=1}^{\mathcal{T}} \prod_{j=1}^{k} (1-q_j).
\end{align*}

\paragraph{Life expectancy in portfolio M}

\begin{figure}[H]
	\centering	\includegraphics[width=.6\textwidth]{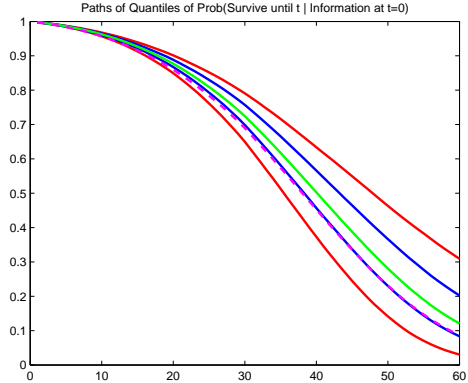}
		\caption{This figure depicts the funnel of doubt for the projected surviving
probabilities of portfolio $M$. The red lines are the estimated 95 \% (resp. 5 \%) yearly quantiles, the lines in blue the 75 \% (resp. 25 \%) yearly quantiles. The green line depicts the b.e. probabilities and the pink dashed line the probabilities derived under the standard formula stress scenario \#4. The graphic is based on 10,000 simulations.}
	\label{fig:funnel_mortality_active}
\end{figure}

We have a remaining life expectancy which is based on the best estimate mortality rates of $\cE_{b.e.} =  38.7$ years. For the stress scenario \#4 the remaining life expectancy decreases to $\cE_{\#4} = 36.9$ years. This means that under the standard formula a reduction of the remaining life expectancy of the portfolio by $1.8$ years over one year is seen as a 1 in 200 year event.\\
In Figure \ref{fig:funnel_mortality_active} we depict the projected probability that the average insured survives the next $t$ years for $t \in [0, \mathcal{T}]$. We see that the stress scenario \#4 corresponds quite precisely to the 25 \% quantile trajectory.

In the next step we estimate how likely such an adjustment by the actuary is, given the new information after one year, using the least-squares Monte Carlo model
\[
	\cE = \alpha + \beta\, AF^{\mathrm{mortality}}(1) + z,
\]
where the error term has conditional expectation zero. The regression output for portfolio $M^{\mathrm{ref}}$ is
\[
	\expectation{\cE \,|\, AF^{\mathrm{mortality}}(1)} \approx \widehat{\cE} = 39.1 - 9.2\,AF^{\mathrm{mortality}}(1),
\]

Given 100,000 simulations, the $0.5\%$ quantile of $\widehat{\cE}$ is $37.7$ years, i.e., a 200-year event would decrease the remaining life expectancy of the portfolio by one year. The stress scenario \#4 postulates a reduction by $1.8$ years that, according to our model, has a probability of near zero.

For the sub-portfolio $M^{\mathrm{sub}}$, the $0.05\%$ quantile of $\widehat{\cE}$ is $37.3$ years, i.e., in a 200-year event one would decrease the remaining life expectancy of the sub-portfolio by $1.4$ years. This clearly stems from the higher basis risk associated with the smaller and less diversified sub-portfolio. Nevertheless, the stress scenario \#4 is still more conservative and has, for the sub-portfolio, a probability of $0.15\%$ and thus a recurrence period above 200 years.


\section{Conclusion}\label{conclusion}

We presented a stochastic framework to quantify the biometric risk of an insurance portfolio that can be used in risk-based solvency regimes. The main difficulty in this context constitutes in the proper representation of long term risks in the profit-loss distribution over a one year horizon. Typically, this requires a mathematically involved or computationally intensive nested simulation approach. In order to resolve this issue, we proposed a least-squares Monte Carlo model to quantify the impact of new experience on the annual re-valuation of the portfolio. Besides being computationally fast and easy to implement, our method shows the contribution of each risk driver to the capital requirement. Since the stochastic model produces the distribution of own funds over one year, any risk measure that is a function of the latter can be easily derived: the Value-at-Risk and Tail-Value-at-Risk used in Solvency II or the Swiss Solvency Test, respectively, constitute two relevant examples.

After developing the necessary theory, we applied the stochastic model to a reference mortality portfolio. The stochastic model proves to be flexible enough to capture the risk inherent in a large and well-diversified portfolio as well as for a smaller sub-portfolio. Of course, special attention has to be paid to a careful parameterization that includes the particular features of the portfolio and adequately reflects mortality risk. The latter has to account for population mortality risk as well as basis risk, i.e., the risk of insureds mortality deviating from population mortality in prospective years.

The aforementioned desirable properties make our proposed model suitable for an internal model under Solvency II. The practical comparison on our reference portfolio shows that our LSMC model provides differentiated results that may lead to a lower (or higher) capital requirement compared to a model based on deterministic stress scenarios like the standard formula. For our reference portfolio, the discrepancy can be attributed to the fact that the pre-defined stresses of the standard formula assume a fixed basis risk and thus implicitly a certain degree of diversification within the portfolio. While this yields similar results for a smaller sub-portfolio of our reference portfolio, the solvency capital requirement of the reference portfolio itself is considerably larger according to the standard formula. 

As a further plausibility check we re-estimated life expectancy of our reference portfolio after one year of additional experience using a linear least-squares Monte Carlo model. The results show reasonable assumption changes for the stochastic model and revealed that the standard formula stress has a recurrence period of well above 200 years for both of our sample portfolios if we condition on the distribution of our stochastic model.

%

\bibliographystyle{plain}
\bibliography{bibliography}

\end{document}